\documentclass[aps,pre,twocolumn,amsmath,amssymb,showpacs,amsfonts]{revtex4}
\usepackage{epsfig}
\usepackage{graphicx}
\usepackage{dcolumn}
\usepackage{bm}

\begin{document}

\title{Inhomogeneous distribution of droplets in cloud turbulence}

\author{Itzhak Fouxon$^{1}$}\email{itzhak8@gmail.com}
\author{Yongnam Park$^2$}
\author{Roei Harduf$^3$} 
\author{Changhoon Lee$^{1,2}$}\email{clee@yonsei.ac.kr}

\affiliation{$^1$ Department of Computational Science and Engineering, Yonsei University, Seoul 120-749, South Korea}
\affiliation{$^2$ Department of Mechanical Engineering, Yonsei University, Seoul 120-749, South Korea}
\affiliation{$^3$ Raymond and Beverly Sackler School of Physics and Astronomy, Tel Aviv University, Ramat Aviv, Tel Aviv 69978, Israel}

\begin{abstract}

We solve the problem of spatial distribution of inertial particles that sediment in turbulent flow with small ratio of acceleration of fluid particles to acceleration of gravity $g$. The particles are driven by linear drag and have arbitrary inertia. The pair-correlation function of concentration obeys a power-law in distance with negative exponent. Divergence at zero signifies singular distribution of particles in space. Independently of particle size the exponent is ratio of integral of energy spectrum of turbulence times the wavenumber to $g$ times numerical factor. We find Lyapunov exponents and confirm predictions by direct numerical simulations of Navier-Stokes turbulence. The predictions include typical case of water droplets in clouds. This significant progress in the study of turbulent transport is possible because strong gravity makes the particle's velocity at a given point unique. 

\end{abstract}


\maketitle

Inhomogeneity of distribution of water droplets in clouds, caused by air turbulence, is significant factor in the formation of rain \cite{Shaw, FFS1, review}. Due to inhomogeneity droplets collide and coalesce more often speeding up the formation of larger drops. It rains when the drops get so large as to reach the ground without evaporating on the way. 

Turbulence-induced inhomogeneities of transported quantities could seem contradicting to the well-known mixing of turbulence producing uniform distribution \cite{Frisch}. Indeed mixing dominates on larger scales. On smaller scales turbulence produces highly irregular spatial structures. Similarly to ordinary centrifuges the rotating turbulent vortices push the inertial droplets out \cite{Maxey} causing very strong non-uniformities in droplets distribution to accumulate with time. Those occur at the typical scale of the vortices (the Kolmogorov scale) which is much smaller than the typical scale of the flow. Thus formation of inhomogeneities of particles by turbulence is small-scale phenomenon, often disregarded due to finite resolution of instruments. Since it is at those scales that droplets collide then the study of the inhomogeneities is necessary to predict rain formation.

Much progress was obtained in the study of non-uniform distributions of inertial particles in the flow when gravity is negligible  \cite{Maxey,BFF,FFS1,review,Bec,Falkovich,BGH,Collins,BecCenciniHillerbranddelta,Stefano,Cencini,Olla,MehligWilkinson,MW,Shaw,Fouxon1,FFS,FP1,FP2,Bewley,caustics,BecRaf}. In the regime where the inertia is not too large, which is where turbulence is most relevant in the rain formation process \cite{review}, the pair-correlation function of concentration was demonstrated to obey a power-law with negative exponent signifying singular (fractal) structure formed by particles in space. Furthermore the leading order term for the exponent at small inertia was obtained for Navier-Stokes turbulence without model-dependent assumptions on the statistics of turbulence \cite{FFS1,Fouxon1}. It depends on statistics of turbulence via the ratio of time integral of correlation function of laplacian of pressure divided by the logarithmic rate of divergence of particles trajectories back in time (third Lyapunov exponent). However those results cannot be applied to droplets in clouds. The latters' motion is influenced by gravity most significantly: the typical value of the ratio $Fr$ of acceleration of air parcels to $g$ is small \cite{review}. Indeed, recent simulations indicate that in this case gravity plays dominant role in the formation of particles' density \cite{Becgr}. Inclusion of gravity into the theory is thus necessary to describe rain.

In this Letter we use the smallness of $Fr$ to derive detailed predictions on the statistics of the spatial distribution of the droplets. These parallel the described predictions for small inertia case but simplify to minimal complexity of dependence on unknown statistics of turbulence: the well-studied energy spectrum. We confirm the predictions by direct numerical simulations of motion of particles in the Navier-Stokes turbulence. 

We demonstrate that similarly to small inertia case the pair correlation function of concentration of not too large droplets whose drag by air is linear (signifying size smaller than $50$ $\mu m$ which is where turbulence is relevant \cite{review}) obeys a power-law with negative exponent. The latter is the ratio of integral of energy spectrum times the wavenumber and $g$ times a numerical coefficient. The exponent is independent of the properties of droplets. Though droplets with different sizes move differently and are distributed instantaneously in different spatial regions, the time or space averaged properties of those distributions are the same universal ones. Universality is stronger: the Lyapunov exponents $\lambda_i$ that describe the long-time deterministic logarithmic rates of growth of lines, surfaces and volumes of the particles obey simple universal relations. The ratio of the principal Lyapunov exponent to the exponent of pair-correlation function is a number dependent neither on droplets, nor on turbulence. The rest of $\lambda_i$ obey similar simple relations. 

This work seems to bring significant progress to extensively studied field \cite{MaxeyRiley,
Maxey,BFF,FFS1,review,Bec,Falkovich,BGH,Collins,BecCenciniHillerbranddelta,Nature2013,fpl,Stefano,Cencini,Olla,MehligWilkinson,MW,Shaw,Seinfeld,Flagan,planetary1,
Engineering1,Engineering2,Biology1,Biology2,Fouxon1,Becgr,Gust,Park2014,FFS,FP1,FP2,Bewley,caustics,BecRaf}. The asymptotic independence of a fractal dimension on $St$ was found in numerical simulations in \cite{Becgr} (see statement of independent work below), see also \cite{Gust}. When gravity is strong, a new pattern of vertical clustering of sedimenting particles was recently observed \cite{Park2014}. However, no theoretical predictions comparable in detail to Eqs.~(\ref{kapyork})-(\ref{spctr}) below that are confirmed numerically were known so far. 

The progress is possible thanks to universality in the distribution of particles in weakly compressible flows \cite{Fouxon1}. 
We demonstrate that gravity has crucial impact on the motion of strongly inertial particles with $St\gtrsim 1$. When gravity is negligible, $Fr\gg 1$, streams of particles ejected from different vortices intersect at the same point where one finds particles with different velocities, the phenomenon sometimes called the sling effect \cite{FFS1,Bewley} or caustics \cite{caustics}. Gravity causes decoherence in the action of turbulent vortices on particles by fast sedimentation through correlated vorticity regions. When $Fr\ll 1$, this decoherence is so significant that the impact of one vortex on particle's motion is negligible - the sedimenting particle leaves the vortex before that catches it to produce the sling, cf. \cite{Becgr,Gust}. It is only smooth accumulated averaged action of many vortices that has finite effect. This causes the particle's velocity to be uniquely determined by its spatial position so that the flow of particles can be introduced where the first order-equation holds,
\begin{eqnarray}&&
\dot {\bm x}(t) =\bm v[t, \bm x(t)],\label{flow}
\end{eqnarray} 
instead of the original second-order classical mechanical equations (Eq.~\ref{basic0} below). The crucial observation is that $\bm v(t, \bm x)$ resulting from complex interplay of inertia and gravity with incompressible driving flow $\bm u(t, \bm x)$ is,  in contrast to $\bm u$, compressible. Thus the particles' density in the steady state is inhomogeneous. 

Similar reduction \cite{Maxey} is well-known in the overdamped limit of strong friction. However, in contrast to that case where $\bm v(t, \bm x)$ can be written explicitly via local spatial and temporal derivatives of $\bm u(t, \bm x)$, in the case of $Fr\ll 1$, $St\gtrsim 1$ the flow,  $\bm v(t, \bm x)$ depends on $\bm u(t, \bm x)$ non-locally so that no explicit formula for $\bm v$ is available. 

To deal with this situation we demonstrate implicitly that $\bm v$ is weakly compressible. This knowledge solely - the existence of the particles' flow and its weak compressibility - implies that particles distribute over fractal set with log-normal statistics that is determined by only one unknown constant - the Kaplan-Yorke codimension $D_{KY}$ that depends on details of velocity statistics \cite{Fouxon1}. 
In particular, that dimension determines the pair-correlation function of concentration $n$ (playing central role in the study of formation of rain)
\begin{eqnarray}&&\!\!\!\!
\langle n(0)n(\bm r)\rangle=\langle n\rangle^2\left(\frac{\eta}{r}\right)^{2D_{KY}},\label{pref}
\end{eqnarray}
where angular brackets stand for spatial averaging. We find $D_{KY}$ in terms of statistics of $\bm u$ not knowing the dependence of $\bm v$ on $\bm u$.  

\begin{figure}
\includegraphics[width=7.0 cm,clip=]{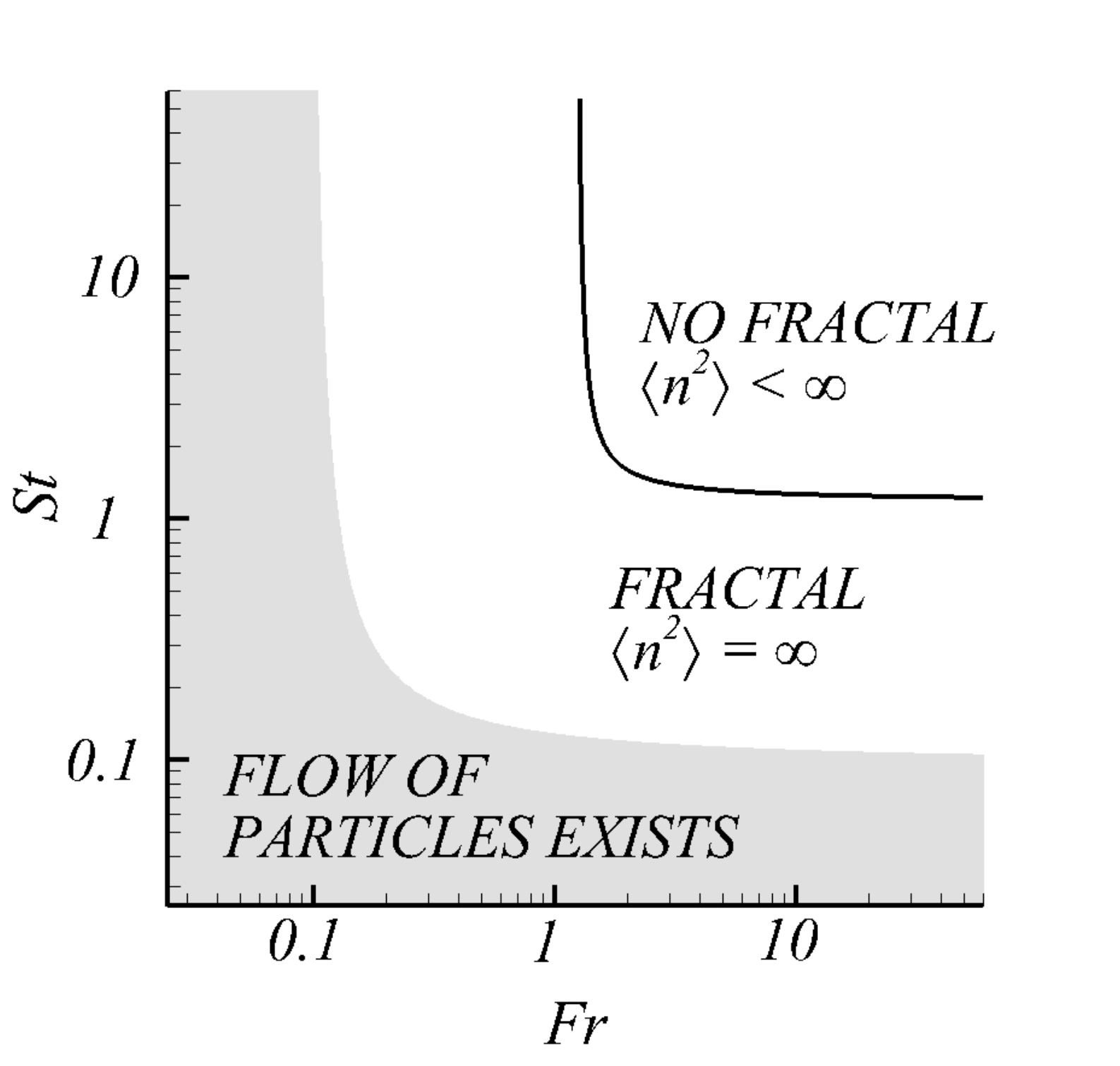}
\caption{In the region $\min[Fr, St]\ll 1$ one can introduce the flow of particles implying fractality of the spatial distribution. The line separating the region of the flow from the region where velocity is significantly multi-valued is not sharp. In contrast, the line separating infinite and finite $\langle n^2\rangle$ is sharp.} \label{fractal}
\end{figure}

It is well-known \cite{Bec,BecRaf} that, in the problem without gravity, there is a transition from fractal singular distribution of particles in space with infinite (due to negativity of the power-law exponent in the pair-correlation function) $\langle n^2\rangle$ at $St<St_{cr}$ to continuous distribution with finite $\langle n^2\rangle$ at $St>St_{cr}$ where $St_{cr}\sim 1$. This implies that at fixed $St>St_{cr}$ the particles' distribution is continuous in the limit of small gravity, $Fr\gg 1$. Since we prove that at $Fr\ll 1$ the distribution is fractal, there is a critical $Fr$ at which the transition from fractal to continuous behavior occurs. This results in the phase diagram in Figure (\ref{fractal}),  cf. \cite{Becgr}. 

We consider small spherical particles with radius $a$ and material density $\rho_p$ driven by incompressible turbulent Navier-Stokes (NS) flow $\bm u(t, \bm x)$ according to
\begin{eqnarray}&&
\ddot {\bm x}(t) =-\left(\dot {\bm x}(t)-\bm u[t, \bm x(t)] \right)/\tau+ \bm g,\label{basic0}\\&&
\partial_t\bm u+(\bm u\cdot \nabla)\bm u=-\nabla p+\nu \nabla^2\bm u,\ \ \nabla\cdot\bm u=0,\label{basic00}
\end{eqnarray}
where $\bm x(t)$ is the particle's coordinate, $p$ is the pressure, $\nu$ is the kinematic viscosity, $\tau=2 \rho_p a^2/9 \nu \rho_f$ is the Stokes relaxation time \cite{MaxeyRiley, Maxey} and the fluid density $\rho_f$ obeys $\rho_f\ll \rho_p$. The flow can be stationary flow sustained by forces (not written explicitly) or quasi-stationary. We assume that one can neglect the particles' interaction and their back reaction on the flow so that each particle obeys Eqs.~(\ref{basic0})-(\ref{basic00}) independently of other particles. 

We study the case of $Fr=\epsilon^{3/4}/[g\nu^{1/4}]\ll 1$, $St\gtrsim 1$ where the impact of gravity is strongest \cite{fpl}. Here the typical acceleration of the fluid particles $\epsilon^{3/4}/\nu^{1/4}$ is written via the energy dissipation rate per unit volume $\epsilon$ (so the Kolmogorov scale $\eta$ is $(\nu^3/\epsilon)^{1/4}$) and the Stokes number $St=\tau\sqrt{\epsilon/\nu}$ is dimensionless inertia of the particle \cite{Frisch}. The consideration holds for droplets in clouds since $Fr \sim 0.01$ for stratocumulus clouds and $Fr \sim 0.06$ for cumulus clouds \cite{review,fpl}. 

The particle drifts through the flow at the velocity $\dot {\bm x}(t)-\bm u[t, \bm x(t)]=\tau \bm g -\tau \ddot {\bm x}(t)$, see Eq.~(\ref{basic0}). Using that acceleration $\ddot {\bm x}(t)$ due to turbulence is close to acceleration of fluid particles at $St\ll 1$ and to acceleration $\sqrt{\epsilon/\tau}$ of eddies with time-scale $\tau$ at $St\gtrsim 1$, one finds \cite{fpl} that the drift is dominated at $Fr=\eta/g\tau_{\eta}^2\ll 1$ by gravity, $\dot {\bm x}(t)-\bm u[t, \bm x(t)]\approx \tau \bm g$. Here $\tau_{\eta}=\sqrt{\nu/\epsilon}$ is the typical time-scale at the Kolmogorov scale so that $\eta/\tau_{\eta}^2$ is the typical acceleration of the fluid particles. The time-scale $\eta/g\tau$ during which the particle traverses $\eta$ to reach uncorrelated regions of the flow is smaller than the Kolmogorov time-scale $\sqrt{\nu/\epsilon}$ so that the timescale of variations of turbulent velocity $\bm u[t, \bm x(t)]$ in the particle reference frame is $\eta/g\tau$. Since velocity gradients are determined by the viscous scale, the sedimenting particle sees gradients $s_{ik}(t)=\nabla_k u_i[t, \bm x(t)]$ change at the time-scale $\eta/g\tau$ as it passes from one correlated region of instantaneous field $\nabla_k u_i(t, \bm x)$ to another. 

If solutions to Eqs.~(\ref{basic0})-(\ref{basic00}) after transients obey Eq.~(\ref{flow}) with certain $\bm v(t, \bm x)$ then $\bm v(t, \bm x)$ obeys the PDE
\begin{equation}\label{eff_vel_cond}
\partial_t \bm v + (\bm v \cdot \bm\nabla) \bm v =
\frac{\bm u - \bm v}{\tau} + \bm g,
\end{equation}
obtained by time differentiation of Eq.~(\ref{flow}) using Eqs.~(\ref{basic0})-(\ref{basic00}), cf. \cite{FFS1,fpl}. The self-consistency demands that $\bm v(t, \bm x)$ evolving according to equation (\ref{eff_vel_cond}) remains well-defined at all times. Indeed, consider initial conditions where particles are distributed in space so that their initial velocity obeys $\bm v(t=0)=\bm v[t=0, \bm x(t=0)]$ where $\bm v(t=0, \bm x)$ is a smooth field. General evolution by Eqs.~(\ref{basic0})-(\ref{basic00}) brings a time $t_*$ when for the first time two particles come to the same spatial point having different velocities, signifying the breakdown of Eq.~(\ref{flow}). This breakdown is signalled by divergence of velocity gradients at $t = t_*$ due to finite difference of $\bm v(t, \bm x)$ at the same point. We conclude that self-consistency of the flow description of solutions to Eqs.~(\ref{basic0})-(\ref{basic00}) demands that there is no finite time blow up of gradients of $\bm v$ obeying Eq.~(\ref{eff_vel_cond}). 

To study the blow up, we observe that $\sigma_{ik}(t, \bm x) \equiv \nabla_k v_i(t, \bm x)$ obey \cite{FFS1}
\begin{equation}\label{vel_gradients1}
\partial_t \sigma + (\bm v \cdot \bm\nabla) \sigma + \sigma^2 =
\frac{s - \sigma}{\tau}.
\end{equation}
In the particle's frame, the gradients $\sigma(t)=\nabla_j v_i[t, \bm x(t)]$ obey the ordinary differential equations (ODE)
\begin{equation}\label{vel_gradients2}
\frac{d \sigma}{dt} + \sigma^2 =-
\frac{\sigma}{\tau}+\frac{s}{\tau}.
\end{equation}
We observe that the gradients $\sigma$ are produced by the gradients $s$ of turbulence in the particle's frame which are finite. If those gradients produce $\sigma\ll 1/\tau$, then the non-linear $\sigma^2$ term is much smaller than the damping term $-\sigma/\tau$ so that the gradients $\sigma$ obey after transients
\begin{equation}\label{linear}
\sigma \approx \sigma_l,
\ \
\sigma_l \equiv \frac{1}{\tau} \int_{-\infty}^{t} s(t') \exp \left( \frac{t'-t}{\tau} \right) dt',
\end{equation}
where the subscript $l$ stands for linear. Clearly in this case $\sigma$ are finite so that the flow description \eqref{flow} is self-consistent. On the contrary, if $s$ produces $\sigma_l\gtrsim 1/\tau$, then the non-linear $\sigma^2$ term in Eq.~(\ref{vel_gradients2}) starts to dominate the dynamics producing a finite-time blow up of $\sigma(t)$ because solutions to $\dot{\sigma}+\sigma^2=0$ blow up in finite time $t_c$ as $(t-t_c)^{-1}$. We conclude that the condition of self-consistency of Eq.~(\ref{flow}) is that 
the probability that $\sigma_l$ is much smaller than $\tau^{-1}$ is close to one, $\left\langle \sigma_l^2 \right\rangle \tau^2\ll 1$. 

One finds from Eq.~(\ref{linear}) that $\sigma_l\sim \int_{t-\tau}^t s(t')dt'/\tau$ so that $\left\langle \sigma_l^2 \right\rangle \sim \tau^{-1} \int_{0}^{\tau} \left\langle s(0)s(t) \right\rangle dt \sim \langle s^2\rangle \eta/g$ where we observed that the correlation time $\eta/g\tau$ of $s(t)$ is much smaller than $\tau$. We find using $\langle s^2\rangle\sim 1/\tau_{\eta}^2$ that $\left\langle \sigma_l^2 \right\rangle \tau^2\sim Fr\ll 1$.  
Thus when $Fr\ll 1$, the solutions to Eqs.~(\ref{basic0})-(\ref{basic00}) are describable by smooth spatial flow.  

Furthermore, the smallness of the correlation time $\eta/g\tau$ of $s$ in comparison with $\tau$ in Eq.~(\ref{linear}) implies that $\sigma\approx \sigma_l$ is Gaussian. Due to isotropy of small scale turbulence that will be presumed below, we have $\langle (\sigma_l)_{ik}\rangle=\langle tr\sigma_l\rangle (\delta_{ik}/3)=0$ where we use that $tr s=0$ by incompressibility of $\bm u$. Thus $\sigma(t)$ is approximately Gaussian noise with zero mean and dispersion $\langle \sigma^2\rangle$ much smaller than the inverse of its correlation time $\tau^2$ (here we note that $\sigma(t)$ is $s(t)$  smoothened over time-scale $\tau$). We conclude that $\sigma(t)$ is short-correlated Gaussian noise. 

We now can find a wealth of predictions on the behavior of particles. Since $\sigma\approx \sigma_l$ and $tr\sigma_l=0$, the flow $\bm v$ is weakly compressible. It was demonstrated in \cite{Fouxon1} that the steady state distribution of particles driven by Eq.~(\ref{flow}) with weakly compressible $\bm v$ has statistics which is completely determined by the Kaplan-Yorke codimension $D_{KY}$, as described in the Introduction. For instance, concentration $n_r$ coarse-grained over scale $r\ll \eta$ obeys log-normal distribution with $\langle n_r^k\rangle=(\eta/r)^{k(k-1)D_{KY}}$, see details in  \cite{Fouxon1,fpl}. 

Thus it remains to find $D_{KY}$ that, for weakly compressible flows, reduces  to $D_{KY}=|\sum\lambda_i|/|\lambda_3|$ \cite{fpl,KY}. Here, $\lambda_i$ are the Lyapunov exponents providing the asymptotic growth rates of logarithms of infinitesimal line, surface and volume elements of particles $l$, $S$ and $V$, respectively. One has $\lim_{t\to\infty} \ln l(t)/t=\lambda_1$, $\lim_{t\to\infty} \ln S(t)/t=\lambda_1+\lambda_2$ and $\lim_{t\to\infty} \ln V(t)/t=\lambda_1+\lambda_2+\lambda_3$,  see \cite{fpl}. To find $\lambda_3$, one can use the results for short-correlated noise \cite{FB,fpl} which give ($2\sigma_l^s=\sigma_l+\sigma_l^t$),
\begin{eqnarray}
&& \lambda_3\approx -\lambda_1\approx \frac{2}{5}\int_{-\infty}^t  \langle tr \sigma_l^s(t) \sigma_l^s(t')\rangle dt'. 
\end{eqnarray}
Using that in the considered limit temporal correlations of $\sigma$ are determined by spatial correlations of velocity gradients, one finds \cite{fpl} 
\begin{eqnarray}&&
|\lambda_3|\approx \lambda_1=\frac{1}{10g\tau}\int_{-\infty}^{\infty} \left\langle \nabla_{k}u_i(0) \nabla_{k}u_i(x) \right\rangle dx.
\label{formula1}
\end{eqnarray}
This can be written via the energy spectrum of turbulence $E(k)$,
\begin{eqnarray}&&\!\!\!\!\!\!\!
|\lambda_3|\tau\approx \lambda_1\tau=\frac{\pi\int_0^{\infty} E(k)kdk}{5 g}\propto Fr,
\label{finalf}
\end{eqnarray}
clarifying the independence of $|\lambda_3|\tau$ of the Stokes number \cite{fpl}. 
Thus instantaneous statistics of turbulence determines the Lyapunov exponent of particles that rapidly traverse the flow that looks to them frozen. In contrast, for passive tracers, different time statistics is relevant.

To find the leading order in $\sigma_l \tau\ll 1$ expression for $tr \sigma$ that determines the rate of growth of volumes $\sum \lambda_i$, we rewrite equation \eqref{vel_gradients2} in the integral form \cite{fpl}
\begin{eqnarray}&&\!\!\!\!\!\!\!\!\!\!\!\!\!\!\!\!
\sigma(t)=\sigma_l(t)
+ \int_{-\infty}^t \exp \left( \frac{t'-t}{\tau} \right) \sigma^2(t') dt'.
\end{eqnarray}
Taking the trace and using $\sigma\approx \sigma_l$, we find \cite{FFS}
\begin{equation}
tr \sigma(t) \approx \int_{-\infty}^t \exp \left( \frac{t'-t}{\tau} \right) tr \sigma_l^2(t') dt'. \label{trace}
\end{equation}
Plugging this into Green-Kubo type formula \cite{FF}
\begin{eqnarray}&&
\sum \lambda_i=-\int_{-\infty}^{0} \langle tr \sigma(0)tr\sigma(t)\rangle dt,
\end{eqnarray} 
we obtain \cite{fpl}
\begin{eqnarray}&&
\sum \lambda_i\!\approx\! -\frac{\tau^2}{2}\int_{-\infty}^{\infty} 
\langle tr \sigma_l^2(0)tr \sigma_l^2(t)\rangle  dt. 
\end{eqnarray}
Finally, using Wick's theorem to find the correlation functions of Gaussian process $\sigma_l(t)$ and performing the time integrals, one finds $\sum \lambda_i\!=-\tau C_{ikpr}C_{kirp}/4$ where 
\begin{eqnarray}&&\!\!\!\!\!\!\!
C_{ikpr}=\int_{-\infty}^{\infty} \langle\nabla_r u_p(0)\nabla_ku_i(\bm g\tau t)\rangle dt,\label{sum2}
\end{eqnarray}
cf. \cite{FouxonHorvai,fpl}. In terms of the energy spectrum, we find 
\begin{eqnarray}&&\!\!\!\!\!\!\!
\sum \lambda_i\tau=-\frac{3[\pi\int E(k)kdk]^2}{32g^2}\propto Fr^2.\label{finalsum}
\end{eqnarray}
Thus volumes of particles decrease exponentially at the rate proportional to $Fr^2/\tau$. The resulting ratio $D_{KY}=|\sum \lambda_i|/|\lambda_3|$ is given by 
\begin{eqnarray}&&\!\!\!\!
D_{KY}=\frac{15\pi \int_0^{\infty}E(k)k dk}{32g}, \label{kapyork}
\end{eqnarray}
where $E(k)$ is the energy spectrum of turbulence. This holds if particles' inertia is not too small so the gravitational distance $g\tau^2$ passed during their relaxation time $\tau$ is much larger than the smallest, Kolmogorov scale of turbulence \cite{Frisch}. Using $\lambda_2=\sum \lambda_i-\lambda_1-\lambda_3$ with correction term relevant at $St\sim 1$ but not $St\gg 1$
\begin{eqnarray}&&\!\!\!\!
\frac{\lambda_1 \tau}{D_{KY}}=-\frac{\lambda_3 \tau}{D_{KY}}=\frac{32}{75},\ \ \lambda_2=-\frac{\lambda_1 D_{KY}}{3},\label{spctr}
\end{eqnarray}
see details in \cite{fpl}. 

\begin{figure}

\includegraphics[width=9.0 cm,clip=]{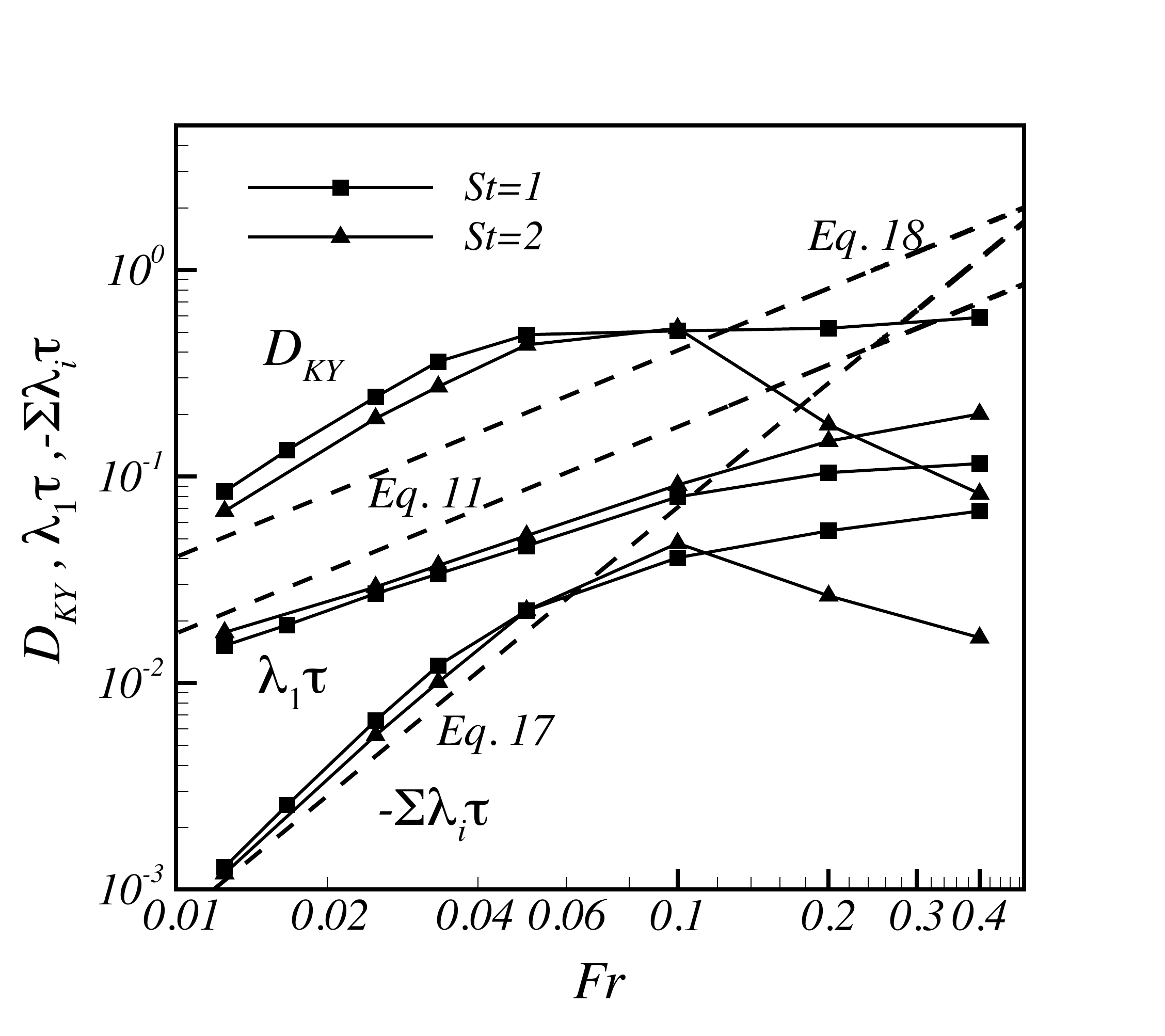}
\caption{Kaplan-Yorke codimension and Lyapunov exponents obtained from direct simulations of particle-laden isotropic turbulence compared to theoretical predictions (dashed lines, Eqs.  \ref{finalf}, \ref{finalsum} and \ref{kapyork} ). Good agreement between numerical results and theoretical predictions is observed when $Fr \leq 0.03$.} \label{Lyapunov}
\end{figure}

We performed numerical simulations to test theoretical predictions for  $\lambda_1 \tau$ (Eq. \ref{finalf}), $\sum \lambda_i \tau$ (Eq. \ref{finalsum}) and $D_{KY}$ (Eq. \ref{kapyork}). Homogeneous isotropic turbulence laden with inertial particles is simulated on a periodic cube. Flow field is obtained from solving the Navier-Stokes equation using a pseudo-spectral method and the particle motion is computed by taking into account the linear Stokes drag and gravity. To resolve the Kolmogorov length-scale fluid motion, $128^3$ grids are used at the Reynolds number based on the the Taylor scale, $Re_{\lambda}=70$. Information of fluid quantities at the particle position is obtained by the fourth-order Hermite interpolation scheme \cite{CYL, LYC}. Details on numerics can be found in \cite{JYL, CKL, AL1, AL2}. The Lyapunov exponents, $\lambda_1$ and $\sum \lambda_i$, are directly computed by releasing many pairs of four particles constructing a tetrahedron. The initial distance between particles in a tetrahedron is set to 1/10,000 of the Kolmogorov length scale and the change of distance between two particles and volume of the tetrahedron is monitored for a period of $45 \tau_{\eta}$ after transient period due to arbitrary initial condition for the particle velocity. 10,000 sets of tetrahedron are released in one flow field and data is collected over a total of 23 flow fields.  Theoretical predictions based on the energy spectrum (Eqs.  \ref{finalf}, \ref{finalsum} and \ref{kapyork}) are compared against numerical results in Fig. \ref{Lyapunov}. As predicted, $\sum \lambda_i$ is negative for small $Fr$ and depends on $Fr$ quadratically as $Fr \to 0$. On the other hand, as $Fr \to 0$, $\lambda_1 \tau$ and $D_{KY}$ depend on $Fr$ linearly, and thus $D_{KY}/(\lambda_1 \tau)$ approaches universal constant. All of them do not show $St$-dependency as $Fr \to 0$, quite distinct behavior compared to no-gravity case \cite{Bec2006}.

The derived universal statistics of particles' attractor (the fractal) at $Fr\ll 1$ is described by one phenomenological constant - $\int E(k) kdk$. This can be rewritten using the spectral viscous scale $\eta_E=(2\nu/\epsilon)\int E(k) kdk$, 
\begin{eqnarray}&&
\eta_E=\frac{\int_0^{\infty} E(k) kdk}{\int_0^{\infty} E(k) k^2dk}=\frac{8\nu}{\pi \epsilon}\int_0^{\infty}\frac{S_2(r)}{r^2}dt,
\end{eqnarray}
where $S_2(r)$ is the second order longitudinal velocity structure function of turbulence \cite{Frisch} and we used $\epsilon=2\nu\int E(k) k^2dk$. The last form stresses that $\eta_E$ is a crossover scale from the viscous to inertial ranges of turbulence \cite{fpl}. 

Our work provides detailed predictions on distribution of water droplets in liquid clouds that are confirmed numerically. No similar predictions were known where previous studies mostly disregarded gravity which impact on droplets' distribution is crucial. The provided formulas can be used directly for studies of formation of rain in wide range of natural situations. 

Further, the results hold for wide range of problems where particles' drag by the flow can be considered linear. Other applications include aerosols spread in the atmosphere \cite{Seinfeld, Flagan}, planetary physics \cite{planetary1}, transport of materials by air or by liquids \cite{Engineering1}, liquid fuel combustion engines \cite{Engineering2}, plankton population dynamics \cite{Biology1, Biology2, Nature2013} and more. 

The separation of particles due to white noise $\sigma$ is different in vertical and horizontal directions \cite{FouxonHorvai}. This is likely to produce a difference in the structure of the fractal in horizontal and vertical directions in accord with \cite{Park2014}. The resulting fractal geometry is the topic of the study in progress \cite{FouxonLee}. 

Our approach can be used to study the behavior of light particles and bubbles as well. This is ongoing work \cite{FouxonLee}.

Finally, we note that the fractal at the scale $l\ll \eta$ forms at the time scale of order $|\lambda_3|^{-1}\ln (\eta/l)$ which is of order $|\lambda_3|^{-1}$ that we demonstrated to be of order of $St/Fr$ times the Kolmogorov time-scale. This time scale is much smaller than the integral time scale of turbulence unless $St/Fr$ is unrealistically large. Thus the described phenomena hold for quasi-stationary turbulence as well. 

When this work was close to finishing, we learnt of the paper \cite{Becgr}. Our results in the questions that were considered in both works are consistent.

This research was supported by a National Research Foundation of Korea (NRF) grant funded by the Korean government (MSIP) (20090093134, 2014R1A2A2A01006544) and Agency for Defence Development.



\end{document}